\documentstyle[aps,multicol,psfig,epsf,epsfig]{revtex}            

\begin{document}
\def\pa{\parallel}
\def\pe{\bot}
\let\a=\alpha \let\b=\beta  \let\c=\chi \let\d=\delta  \let\e=\varepsilon
\let\f=\varphi \let\g=\gamma \let\h=\eta \let\k=\kappa  \let\l=\lambda
\let\m=\mu   \let\n=\nu   \let\o=\omega    \let\p=\pi
\let\r=\varrho  \let\s=\sigma \let\t=\tau   \let\th=\vartheta
\let\y=\upsilon \let\x=\xi \let\z=\zeta
\let\D=\Delta \let\F=\Phi  \let\G=\Gamma  \let\L=\Lambda \let\Th=\Theta
\let\O=\Omega       
\draft
\tightenlines

\begin{multicols}{2}
\narrowtext

{\bf Albano and Saracco reply.} In our paper \cite{we}, it is assumed
that, at criticality and starting form a ground-state configuration, 
the order parameter ($OP$) decays according to   
\begin{equation}
OP(t) \propto  t^{-\beta/\nu_{\pe} z},
\label{orpa}
\end{equation}
\noindent (Eq. (9) in \cite{we}), where it is implicit that $z = z_{\pa}$.
In the preceding comment, Caracciolo et al. (CGGP) \cite{tanos} have
suggested, instead, that $z$ must be interpreted as $z_{\pe}$. Based
on this assumption CGGP conclude that the exponent $\beta/\nu_{\pe} z$
reported in \cite{we} supports the field-theoretical equation developed 
in reference \cite{JS} (Eq.(1) in \cite{we}), while the numerical 
estimate of the exponent $c_{\pe}$ differs from the prediction of both
field-theoretical models considered by us, given by Eqs.(1) and (2)
in \cite{we}, which were taken from references \cite{JS} and \cite{gallegos},
respectively. The conclusions of CGGP are in contrast to the main
finding of our paper that fully supports the universality class 
predicted by Eq.(2) in \cite{we}.

Let us now show that the agreement between the exponent 
$\beta/\nu_{\pe} z_{\pe}$ predicted by Eq.(1) in \cite{we} and our numerical 
data is merely coincidental and that the conclusions of CGGM are 
inconsistent since the relevance of the exponent $c_{\pe}$ to determine the
universality class is simply disregarded.

Taking the logarithmic derivative of Eq.(\ref{orpa}) at criticality one has
\begin{equation}
\partial{ln} OP(t,\tau) |_{\tau = 0} \equiv OP* \propto t^{1/\nu_{\pe}z}.
\label{deri}
\end{equation}
\noindent Furthermore, inserting Eq.(\ref{deri}) in Eq.(\ref{orpa}) one gets
$ln(OP) \propto -\beta ln(OP*)$, which gives an unbiased estimation 
of $\beta$ independent of any assumption about $z$. Our data 
shown in Figure 1 are well fitted by an exponent $\beta = 0.330$, in 
agreement with the universality class predicted by Eq.(2) in \cite{we},
while the value $\beta = 1/2$ predicted by Eq.(1) in \cite{we} can clearly 
be ruled out.
\begin{figure}
\centerline{
\epsfxsize=8cm
\epsfysize=6cm
\epsfbox{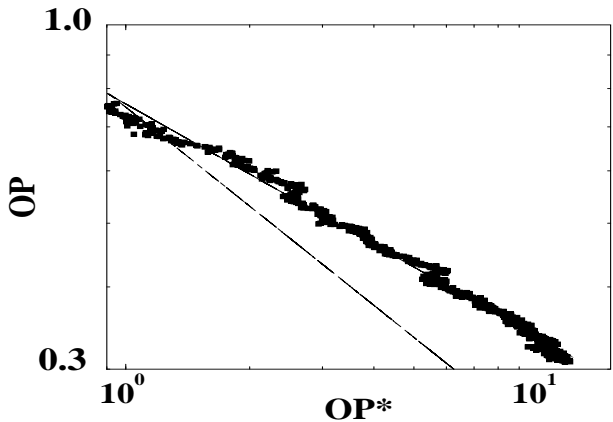}}
\caption{Log-log plot of $OP$ {\it versus} the logarithmic
derivative of $OP$. The full line with slope $\beta = -0.330$
is the best fit of the data. The dashed line has slope $\beta = 1/2$.}
\label{Figure1}
\end{figure} 

>From Eq.(\ref{deri}) it also follows that 
$OP*^{\nu_{\pe}} \propto t^{1/z}$. Our numerical results shown in figure 
2 are in full agreement with $z = z_{\pa} \simeq 1.998$, while
the value $z_{\pe} \simeq 4$ suggested by CGGM can clearly be ruled out.

\begin{figure}
\centerline{
\epsfxsize=8cm
\epsfysize=6cm
\epsfbox{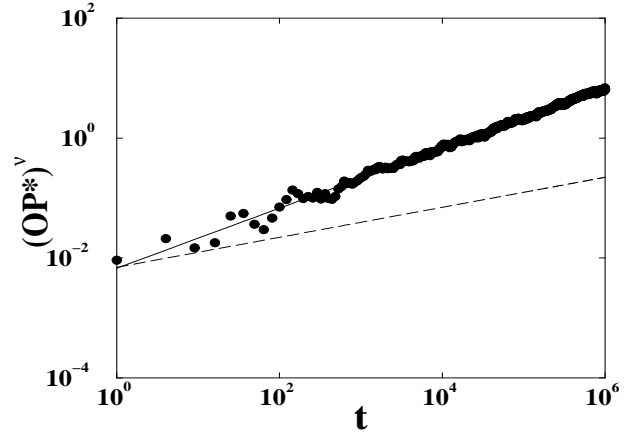}}
\caption{Log-log plots of the $OP*^{\nu_{\pe}}$ {\it versus} $t$.  
The full line with slope $1/z_{\pa} = 0.501$ is the best fit of the data. 
The dashed line has slope $1/z_{\pe} = 0.25$.}
\label{Figure2}
\end{figure}

Summing up, we have provide conclusive evidence showing that,
in contrast to the suggestions of the preceding comment \cite{tanos}:
1) The decay of the order parameter is governed by the time evolution 
of the correlation length parallel to the field and consequently the exponent 
$z$ of Eq.(1) must be identified as $z_{\pa}$ as in our original paper 
\cite{we}; 2) A numerical determination of the exponent $\beta$ can
be performed independently of any assumption on the value of $z$;
3) All exponents measured are fully consistent with the universality 
class predicted by Eq.(2) in \cite{we} and developed in 
reference \cite{gallegos}.

{\bf Acknowledgments}: This work was supported by CONICET, 
UNLP and ANPCyT (Argentina). G.S. acknowledges the CIC for 
a research fellowship.

Ezequiel V. Albano and Gustavo Saracco.
Instituto de Investigaciones Fisicoqu\'{\i}micas Te\'oricas y 
Aplicadas (INIFTA), UNLP, CONICET, Casilla de Correo, 
16 Sucursal 4, (1900) La Plata, Argentina.

\end{multicols}

\begin{thebibliography}{99}
\vspace*{-1.75cm}

\bibitem{we} E. V. Albano and G. Saracco. Phys. Rev. Lett, 
{\bf 88} 145701 (2002). 

\bibitem{tanos} S. Caracciolo, A. Gambassi, M. Gubinelli and A.  
Pelissetto, preceeding comment,. Phys. Rev. Lett. {\bf XX}, xxxxxx (2003).

\bibitem{JS} H. K. Janssen and B. Schmittmann, Z. Phys. B {\bf 64}. 503
(1986). K.-t. Leung and J. L. Cardy, J. Stat. Phys. {\bf 44}, 567
(1986); ibid {\bf 45}, 1087 (Erratum) (1986).

\bibitem{gallegos} P. L. Garrido, F. de los Santos and M. A. Mu\~noz,
Phys. Rev. E {\bf 57}, 752 (1998), and  P. L. Garrido, M. A. Mu\~noz, 
and F. de los Santos, Phys. Rev. E {\bf 61}, R4683 (2000). 
 
\end{thebibliography}
\end{document}